\documentclass[preprint,aps,prl,showpacs,preprintnumbers,amsmath,amssymb,superscriptaddress]{revtex4-1}

\usepackage{graphicx}% Include figure files
\usepackage{dcolumn}% Align table columns on decimal point
\usepackage{bm}% bold math
\usepackage{bibtopic}
\usepackage{multirow}
\usepackage{subfigure}
\usepackage{xcolor}
\usepackage[normalem]{ulem}
\usepackage{soul}
\begin{document}

%\preprint{AIP}

\title{The driving force for charge ordering in rare earth nickelates}

      \author{Basudeb Mandal}

      \affiliation{Department of Condensed Matter Physics and Material Sciences, S N Bose National Centre for Basic Sciences, Block-JD, Sector-III, Salt Lake, Kolkata - 700106, West Bengal, India.}

      \author{Sagar Sarkar}

      \affiliation{Department of Condensed Matter Physics and Material Sciences, S N Bose National Centre for Basic Sciences, Block-JD, Sector-III, Salt Lake, Kolkata - 700106, West Bengal, India.}

            \author{Shishir Kumar Pandey}

      \affiliation{Department of Condensed Matter Physics and Material Sciences, S N Bose National Centre for Basic Sciences, Block-JD, Sector-III, Salt Lake, Kolkata - 700106, West Bengal, India.}

            \author{Priya Mahadevan}

      \affiliation{Department of Condensed Matter Physics and Material Sciences, S N Bose National Centre for Basic Sciences, Block-JD, Sector-III, Salt Lake, Kolkata - 700106, West Bengal, India.}

                  \author{Cesare Franchini}

      \affiliation{University of Vienna, Faculty of Physics and Center for Computational Materials Science, Vienna 1090, Austria}

                  \author{A. J. Millis}

      \affiliation{Department of Physics, Columbia University, New York, New York 10027, USA}

                        \author{D. D. Sarma}

      \affiliation{Solid State and Structural Chemistry Unit, Indian Institute of Science, Bangalore, India}
     \email[E-mail:]{priya.mahadevan@gmail.com,sarma.dd@gmail.com}

\begin{abstract}

%An almost ubiquitous charge ordering has been observed among the rare-earth(RE) nickelates of the RENiO$_3$ with the exception of LaNiO$_3$ which remains metallic. In this work we introduce an additional potential on the Ni d states and vary the charge transfer energy ($\Delta$) between the O p and Ni d states. The onset of  holes on the oxygen is dictated by $\Delta_{eff}$ = $\Delta$ - $\dfrac{W}{2}$ where $W$ is the bandwidth of the oxygen p band. Negative values of this parameter we find dictates the presence of charge ordering which is quantified in terms of inequivalent Ni-O bondlengths and magnetic moments for the two Ni atoms.
We show that charge ordering (more precisely, two-sublattice bond disproportionation) in the rare earth nickelate perovskites is intimately related to a negative charge transfer energy. By adding an additional potential on the Ni d states we are able to vary the charge tranfer energy and compute relaxed structures within an ab-initio framework. We show that the difference in Ni-O bond lengths and the value of the ordered state magnetic moment correlate with the charge transfer energy and that the transition to the bond-disproportionated state occurs when the effective charge transfer energy becomes negative.
\end{abstract}

\maketitle

Although the 3d transition metal oxides have been studied since the 1950's,  improved growth and characterization techniques as well as new theoretical approaches have continued to yield new insights.\cite{MKWu_PRL_1987,RVHelmolt_PRL_1993,AAsamitsu_Nature_1995,JChakhalian_nature_2013,SStemmer_SciRep_2016}. The rare earth perovskite nickelates are of particular current interest. These materials  exhibit metal-insulator transitions for all members of the family RENiO$_3$ (where RE denotes a rare earth ion), with the exception of RE = La\cite{Catalan_PhaTran_2008,Middey_ARMR_2016}. The metal-insulator transition is coincident with a crystal distortion in which the mean Ni-O bond length alternates between  two inequivalent Ni sites, defining a bond disproportionation. \cite{JAAlonso_PRL_1999,JAAlonso_PRB_1999,JAAlonso_JACS_1999,UStaub_PRL_2002}.  This state is sometimes also referred to as ``charge ordered".%The presence of the rare earth atom in the perovskite lattice has always been understood as controlling the structural distortions. An atom with a smaller ionic radius leads to a smaller volume of the unit cell. However, this would also imply shorter Ni - O bondlengths. This shortening of the bondlengths has the effect of increasing the Coulomb repulsion between electrons on Ni and those on oxygen. The NiO$_6$ octahedra rotate leading to longer Ni-O bondlengths. This results in Ni -O- Ni  angles which deviate from 180$^o$, with the smaller RE ion resulting in larger deviations of the Ni-O-Ni angle. The bandwidth of the Ni d states is controlled by the effective hopping interaction strength, which depends on the Ni-O-Ni angle.  Initially the metal-insulator transition in the nickelates was understood as being driven by the modified bandwidth arising from the rotation of the NiO$_6$ octahedra\cite{Sarma_PRB_date}. However, later analysis of the structure revealed a breathing mode distortion associated with the NiO$_6$ octahedra. One Ni atom had an expanded NiO$_6$ octahedron associated with it, while the other had a contracted NiO$_6$ octahedron associated with it\cite{JAAlonso_PRL_1999}. While there was no significant charge difference between the two Ni sites, the associated Ni-O bondlengths led to one of the Ni atoms with longer Ni-O bondlengths ($\sim$ 2\AA) being labelled Ni$^{2+}$, while the other with shorter Ni-O bondlengths ($\sim$ 1.9\AA) was labelled Ni$^{4+}$. A similar finding has emerged in the context of other charge ordered nickelates\cite{JAAlonso_PRL_1999,JAAlonso_PRB_1999,JAAlonso_JACS_1999,UStaub_PRL_2002}.

While the rare earth perovskite nickelates exhibit bond disproportionation, the rare earth perovskite cobaltates  formed with the neighbouring transition metal atom Co in the same oxidation state exhibit no such ordering. An important parameter that controls the electronic structure for the late transition metal oxides is the charge transfer energy, given by the energy required to transfer an electron from the oxygen p levels to the transition metal d levels. The charge transfer energy decreases as one goes across the 3d transition metal series from Ti to Cu\cite{Mahadevan_PRB_1997}  and it is natural to associate the change in charge transfer energy with the propensity to bond disproportionation.

Formal valence considerations assign the d$^7$ configuration to the Ni in the RENiO$_3$ perovskites. However, if the charge transfer energy is strongly negative, the electronic configuration is more appropriately represented as d$^8$$\bar{L}$ (with the $\bar{L}$ denoting a hole on the ligand).  The importance of an effectively negative charge transfer energy in this family of compounds was first pointed out by Barman et al.\cite{Sarma_PRB_1994} while discussing the insulating ground state of NdNiO$_3$ in contrast to the metallic one of LaNiO$_3$. Mizokawa et al.\cite{TMizokawa_PRB_2000} carried out model Hamiltonian calculations for a multiband Hubbard model and could capture the bond disproportionation at a negative value of the charge transfer energy when they included a breathing mode distortion of the NiO$_6$ octahedra. This suggests that the combination of  lattice distortions and a negative charge transfer energy drove the charge ordering. Mazin and coworkers\cite{Mazin_PRL_2007} argued that %the charge ordering was an alternative to Jahn-Teller distortions, and
part of the energy lowering associated with the disproportionation  came from the energy gain from Hund's intra atomic exchange interactions, which favor a high-spin d$^8$ state.  Building on the Mizokawa picture, Park, Millis and Marianetti \cite{HPark_PRL_2012} presented density functional plus dynamical mean field calculations that explained the disproporation in terms of a site selective Mott transition occurring in a situation in which the charge transfer energy was very negative, and Johnson and collaborators later considered the same physics in a model system perspective \cite{Sawatzky_PRL_2014}. On the other hand, Peil and Georges \cite{Subedi14} argued that an appropriate low energy description of the physics was in terms of a Hubbard model with a vanishing or negative U; in this effective low energy picture the bond-disproportionated state is indeed characterized by charge order.

In this paper we take a new approach  to this issue by examining in more detail the connection between bond disproportionation and the charge transfer energy. Introducing a potential on the Ni d states, we are able to vary the charge transfer energy and examine the ensuing changes in the structure as well as the electronic structure within an {\it ab initio} framework in contrast to all model Hamiltonian approaches in the past. We find that the onset of charge ordering is characterised by the point at which the Ni $d$ band enters the oxygen $p$ band, defining the effective negative charge transfer energy\cite{Sarma JSSC 1989, Nimkar PRB 1993}. This destabilizes the RE-oxygen network which is otherwise ionic,  driving the charge ordering.

The electronic structure of NdNiO$_3$ was calculated within a projected augmented wave\cite{PEBlochl_PRB_1994} implementation of density functional theory within the Vienna Ab initio simulation package (VASP)\cite{GKresse_PRB_1996,GKresse_CompMatSci_1996} code. The experimental lattice parameters were taken\cite{GarciaMunoz_PRB_2009}. The magnetic structure (both T-AFM, non-collinear E$^{\prime}$-AFM and FM) was imposed and the electronic structure was calculated within the Dudarev implementation\cite{Dudarev_PRB_1998} of GGA+U with a U of 4 eV on the Ni sites. The generalized gradient approximation\cite{JPPerdew_PRL_1997} was used for the exchange correlation functional. A k-points grid of 4x6x2 was used for calculating the electronic structure. While the lattice parameters were kept fixed at the experimental values, the internal positions were optimised to find the minimum energy configuration so that the forces were less than 10$^{-3}$ eV/$\AA$. The general features of the structure are similar when we assume ferromagnetic order. Consequently the rest of the analysis in terms of microscopic model has been carried out for the ferromagnetic unit cell which is smaller. A k-point mesh of 6x4x6 and an energy cutoff of 500 eV was used. Spheres of radii 1 $\AA$ are constructed around each atom for the calculation of the density of states and magnetic moment and within the spheres centered on the Ni ions  a d-symmetry potential of constant radial part is introduced.  The structure is then optimised to find the structural and magnetic parameters in the presence of the potential and the charge transfer energy is quantified by using maximally localized wannier function methods\cite{Mostofi_CPC_2008,Marzari_PRB_1997,Souza_PRB_2001} to  map the ab-initio band structure onto a tight binding model using the VASP to Wannier90 interface \cite{Franchini_JPCM_2012}.  The results are used to construct a schematic diagram of the electronic structure.

There are two candidate orderings which have been proposed for the magnetic structure of the magnetic nickelates. The first corresponds to an up-up-down-down ordering of the spins on the Ni along the three pseudo cubic directions, and has been referred to as T-AFM type magnetic structure\cite{Garcia_PRB_1994}. There are variants that differ slightly in the stacking of these chains, and differ slightly in the total energy\cite{Lezaic_PRB_2013}. The other structure corresponds to a non-collinear one in which the neighbouring spins have equal magnitude but are rotated by 90$^o$. We have used  both of these structures to initialize our calculations;  we find that  both cases relax  to the same magnetic configuration.  The fully relaxed structure contains Ni-O$_6$ octahedra of short mean bond length ($\sim$ 1.90  $\AA$ for the case with no extra Ni potential applied)  and NiO$_6$ octahedra of longer mean bond length ($\sim$ 2.0  $\AA$ if no Ni potential is applied). The Ni sites with short-bond octahedra have a zero magnetic moment, while the Ni sites with long-bond octahedra  have a magnetic moment of 1.50 $\mu_B$  (Ni$^{2+}$). A similar difference of moment was found experimentally and was initially interpreted as a Ni charge disproportionation \cite{JAAlonso_PRL_1999}.  However, examining the density of states associated with each of the Ni sites (Fig. 1), we find that the $t_{2g}$ states on both Ni sites are completely filled, while the mean occupancy of the $e_g$ states on both sites is $\sim$ 2. On the long-bond Ni site the majority spin $e_g$ channel is found deep inside the valence band and is fully occupied, while the minority spin  $e_g$ channel is empty, with a very small admixture of O $p$ implying an Ni $d^8$ configuration. On the short-bond Ni sites the high-lying $e_g$ states are found at $\sim$ 1-2 eV in the conduction band and have significant O $p$ admixture; these are antibonding states; the corresponding bonding states are located deep inside the valence band. The significant O $p$ admixture suggests that one should associate an electronic configuration of $d^8 \underline{L}^2$ as previously suggested\cite{TMizokawa_PRB_2000,HPark_PRL_2012,Sawatzky_PRL_2014}. As discussed by Park et al \cite{HPark_PRL_2012} the  spin splitting of these states is very small (zero in the present calculation), indicating that the holes on the oxygen states form a singlet with the Ni $e_g$ electrons which accounts for the zero magnetic moment.
\begin{figure}[h]
\includegraphics[angle=0,width=11.25cm,height=12.75cm]{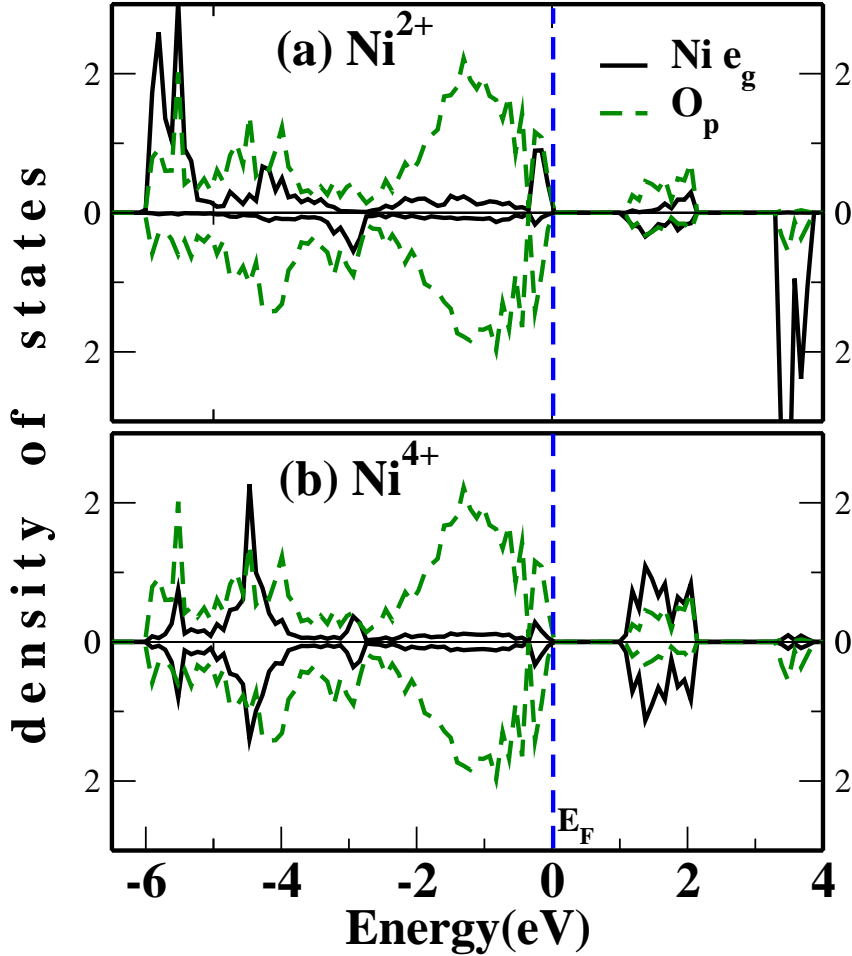}
\caption{The spin resolved (upper and lower panel) Ni $e_g$ and O $p$ contributions to the density of states for (a) Ni$^{2+}$, (b) Ni$^{4+}$ sites in
NdNiO$_3$ considering the T-type antiferromagnetic structure and U=4~eV on Ni.}
\label{FIG.1}
\end{figure}

\begin{figure}[h]
\includegraphics[angle=0,width=9.92cm,height=6.76cm]{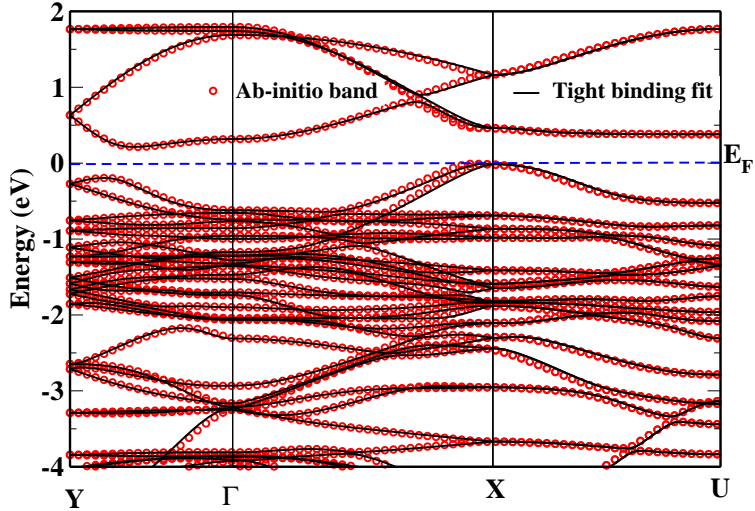}
\caption{A comparison of the ab-initio band structure and the tight binding fit for ferromagnetic NdNiO$_3$ at a $\Delta$ = -0.87 eV}
\label{FIG.2}
\end{figure}

Having established that the DFT+U calculations correctly reproduce the basic physics of NdNiO$_3$, we analyse the consequences of varying the charge transfer energy. For convenience in the analysis and interpretation we consider a ferromagnetic ground state (which can also be stabilized in the DFT+U method, although it is not the true ground state). In the ferromagnetic state the inequivalent Ni sites have respectively a large and a small moment, but in contrast to the T-antiferro state the smaller moment, while much less than the larger one, is not  zero.  We vary the potential acting on the Ni, and for each value of the potential determine  the on-site magnetic moments, the amplitude of the bond disproportionation, and the charge transfer energy as defined from the Wannier fit. % prIt has also been seen in the past that a fitting which includes the states contributing in the energy window allows a discussion of the trends in the electronic structure. This gives us confidence in a discussion in the trends of the extracted energies. In the rest of the discussion, we measure $\Delta$ as the difference in onsite energies between the majority spin 2+ Ni levels with e$_g$ symmetry and the oxygen p levels.

Because the charge transfer energy is a monotonic function of the on-site potential, we plot the magnetic moments and mean octrahedral bond lengths against charge transfer energy $\Delta$ in Fig. 3.% In contrast to the result for the T-type anti-ferro magnetic configuration where the Ni$^{4+}$ sites had a zero magnetic moment associated with them, in the ferromagnetic configuration we find that the Ni$^{4+}$ sites have a finite magnetic moment associated with them. In Fig. 3(b) we have plotted the variation in the magnetic moment as a function of $\Delta$. We find a magnetic moment of $\sim$ 1.45 $\mu_B$ for negative values of $\Delta$ which almost remains invariant as $\Delta$ is increased and even becomes positive. There is an increase in the moment from 0.5  $\mu_B$ for the Ni$^{4+}$ sites. It gradually approaches the magnetic moment associated with Ni$^{2+}$ sites and the two become equal at some value of positive $\Delta$. The existence of the two Ni sites with different magnetic moments may be regarded as a  signature of the presence of charge ordering. These results suggests that the charge ordered state exists for negative $\Delta$, and the state becomes less stable as $\Delta$ becomes more positive. While these results suggests that $\Delta$ is indeed playing a role, its presence at positive values of $\Delta$ is a bit puzzling.
As $\Delta$ is increased from the value  $\approx -1.15~eV$, Fig. 3a shows that the mean bond length of the short-bond octahedra increases, while the mean bond length of the long-bond octahedra changes only slightly. For charge transfer energies greater than about $1~eV$ the difference between the two mean octahedral bond lengths becomes negligible. Fig 3b shows a similar increase in the magnetic moment of the short-bond site as the charge transfer energy is increased, with the difference in moments between sites becoming negligible for $\Delta \gtrsim 1~eV$.

\begin{figure}[h]
\includegraphics[angle=0,width=10cm,height=10cm]{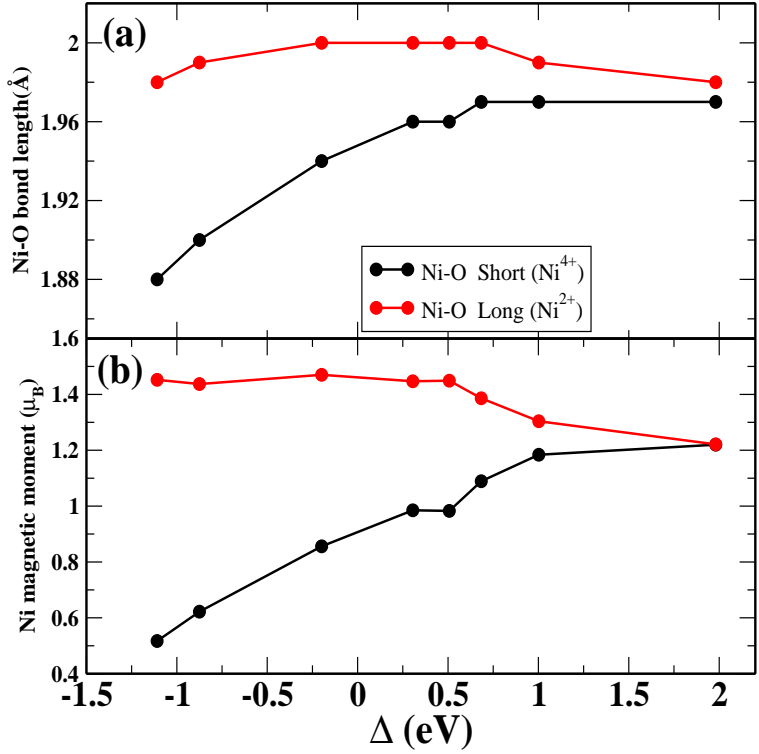}
\caption{Variation in the (a) Ni-O bondlengths and (b) Magnetic moments on the Ni sites with $\Delta$ }
\label{FIG.3}
\end{figure}

%The local environment of the Ni atoms also showed changes as $\Delta$ was varied. The Ni$^{2+}$ bondlengths varied slightly from 1.98 to 2.0 $\AA$ as $\Delta$ was increased. However, the Ni$^{4+}$ bondlengths showed larger variations. At negative values of $\Delta$, Ni$^{4+}$ bondlengths were $\sim$ 1.88 $\AA$ and  these increased gradually to 2.0 $\AA$. At the same point that we had the same magnetic moment on both the Ni sites, we also had the same local environment.

\begin{figure}[h]
\includegraphics[angle=0,width=10cm,height=12cm]{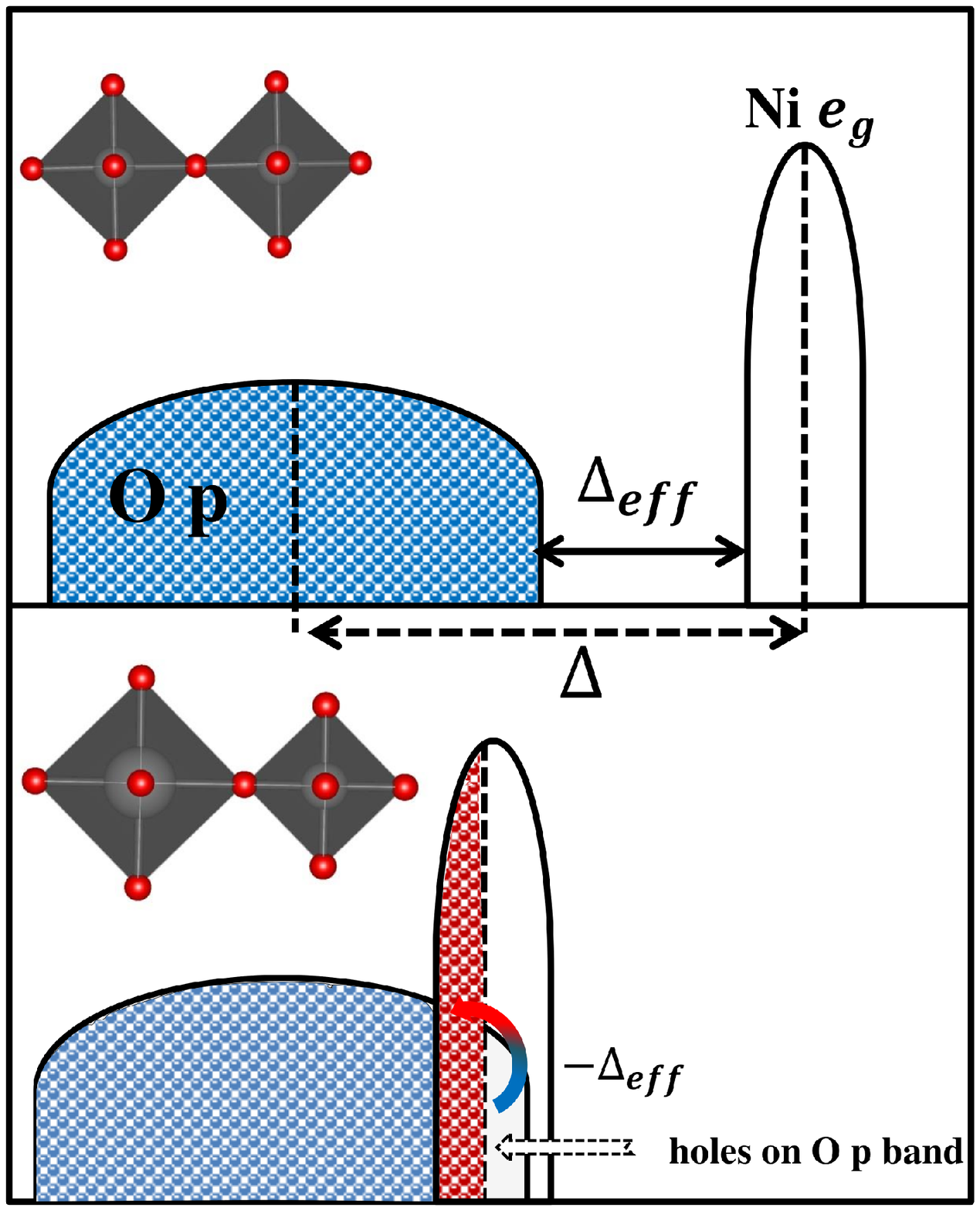}
\caption{Schematic indicating the definitions $\Delta$ and $\Delta_eff$ used in the text. While a positive $\Delta_{eff}$ has uniform
NiO$_6$ octahedra, a negative $\Delta_{eff}$ leads to transfer of holes to the oxygen p bands and occurrence of a breathing distortioni of the NiO$_6$ octahedra. }
\label{FIG.4}
\end{figure}

Having established that the charge transfer energy controls the disproportionation physics, we now consider in more detail the mechanism. We have calculated the width of the oxygen $p$-band within the tight-binding model by switching off the $p$-$d$ interactions. %In order to understand the value of $\Delta$ at which we had the onset of charge ordering, we calculated the bandwidth of the oxygen p band. This was done by switching off the p-d interactions and calculating the width of the oxygen p band. This is found to be around 5.5 eV. Hence the onset of  charge ordering seems to be associated with the point at which the holes begin to occupy the oxygen p band.  %At a qualitative level this can be understood as follows. In contrast to the transition metal - oxygen network which is very covalent, the rare-earth - oxygen network is very ionic. With the holes moving onto the oxygens, this ionic network is destabilized. Beyond a critical number of holes one finds that the structure distorts, driving the charge ordering instability.
We find that the disproportionation disappears when the charge transfer energy become large enough that the $p$-band becomes  filled, as shown schematically in Fig. 4. This supports the view \cite{HPark_PRL_2012} that the disproportionation arises from a preferential hybridization of the ligand holes with one of the Ni states.

Future work will examine trends across the nickelate series from this point of view, and will consider the relation between the charge transfer energy and the effective U of the low energy theory introduced by Subedi, Peil and Georges \cite{Subedi14}.

%The origin of charge ordering in the rare earth nickelates has been examined within first principle electronic structure calculations  for the case of NdNiO$_3$. The charge transfer energy between the oxygen $p$ and the transition metal $d$ states has been systematically varied. The onset of charge ordering is found to be associated with the point when holes are located on the oxygen $p$ band, and is shown to take place beyond a critical concentration.

PM,AJM and DDS thank the Indo-US Science and Technology Forum for a Joint Network Centre under which part of the project was carried out. PM, CF and DDS thank DST and FWF for a joint project. SS thanks CSIR, Government of India for a fellowship. AJM acknowledges the support of the Basic Energy Sciences Division of the US Department of Energy under grant ER-046169.

\bibliography{References}

\end{document}